\documentclass[12pt]{iopart}
\usepackage[latin1]{inputenc}
\usepackage{amsfonts}
\usepackage{amssymb}
\usepackage{graphicx}
\usepackage{enumerate}

\begin{document}

\title[Probing photo-ionization: Simulations $\ldots$]{Probing photo-ionization: Simulations of positive streamers in varying N$_2$:O$_2$-mixtures}

\author{G Wormeester$^{1}$, S Pancheshnyi$^{2}$, A Luque$^{3}$, S Nijdam$^{4}$, U Ebert$^{1,4}$}
\address{$^{1}$ Centrum Wiskunde \& Informatica (CWI), PO Box 94097, 1090 GB Amsterdam, The Netherlands}
\address{$^{2}$ Laboratoire Plasma et Conversion d'Energie, 118 Route de Narbonne, 31062 Toulouse cedex 4, France}
\address{$^{3}$ Instituto de Astrof\'{i}sica de Andaluc\'{i}a, IAA-CSIC, PO Box 3004, 18080 Granada, Spain}
\address{$^{4}$ Eindhoven University of Technology, Department of Applied Physics, PO Box 513, 5600 MB Eindhoven, The Netherlands}
\eads{G.Wormeester@cwi.nl}

\begin{abstract}
Photo-ionization is the accepted mechanism for the propagation of positive streamers in air though the parameters are not very well known; the efficiency of this mechanism largely depends on the presence of both nitrogen and oxygen. But experiments show that streamer propagation is amazingly robust against changes of the gas composition; even for pure nitrogen with impurity levels below 1 ppm streamers propagate essentially with the same velocity as in air, but their minimal diameter is smaller, and they branch more frequently. Additionally, they move more in a zigzag fashion and sometimes exhibit a feathery structure. In our simulations, we test the relative importance of photo-ionization and of the background ionization from pulsed repetitive discharges, in air as well as in nitrogen with 1~ppm O$_2$. We also test reasonable parameter changes of the photo-ionization model. We find that photo-ionization dominates streamer propagation in air for repetition frequencies of at least 1~kHz, while in nitrogen with 1~ppm O$_2$ the effect of the repetition frequency has to be included above 1~Hz. Finally, we explain the feather-like structures around streamer channels that are observed in experiments in nitrogen with high purity, but not in air.
\end{abstract}

\submitto{\JPD -- 19 August 2010, revised 21 September 2010 and 21 October 2010}

\maketitle
\section{Introduction}

\subsection{Positive streamers in varying gases}

Streamers are thin channels of ionized gas that appear when a high voltage is applied to a large gas volume \cite{Raizer:1991,Veldhuizen:2000,Ebert:2006,Ebert:2008};  they are of significant importance in atmospheric electricity (for example, in lightning and sprites \cite{Pasko:2007,Ebert:2010}) as well as in industrial applications such as lighting, removal of volatile organic components and disinfection \cite{Heesch:2008} and plasma-assisted combustion \cite{Starikovskaia:2006}. We distinguish between positive and negative streamers, where positive streamers carry a net positive charge at their heads and propagate in the direction of the ambient electric field, while negative streamers carry a negative head charge and propagate against the field. This means that negative streamers propagate in the direction of the electron drift, while positive streamers move against the electron drift direction; therefore they require a source of electrons ahead of the streamer to support the impact ionization process and the further growth of the ionized area at the streamer head. Despite the fact that positive streamers propagate against the electron drift velocity, in air they appear more easily than negative streamers and they propagate faster. This faster propagation was observed in experiments \cite{Briels:2008} and explained in \cite{Luque:2008}: in negative streamers, the electrons at the side of the streamer channel drift outwards and reduce the field focussing at the streamer tip while positive streamers stay narrow and therefore enhance the electric field at the streamer tip to higher values.

While most work focusses on positive streamers in air, positive streamers have been observed in varying nitrogen:oxygen ratios \cite{Yi:2002,Briels:2008c,Briels:2008b,Nijdam:2010}, in argon \cite{Aleksandrov:2001} as well as in gas mixtures presenting the atmospheres of Venus (CO$_2$:N$_2$) and of Jupiter-like gas giants (H$_2$:He) \cite{Dubrovin:2010}. In a recent experiment \cite{Nijdam:2010}, positive streamers were observed in nitrogen, oxygen and argon with impurity levels below 1 ppm; they propagate with essentially the same velocity as in air, but are thinner and less straight, they branch more, move more in a zigzag fashion or even can form feathery structures.

The traditional explanation for positive streamer propagation in air is photo-ionization; a review of the history of the concept can be found in~\cite{Nijdam:2010}. But photo-ionization according to the traditional model critically depends on the ratio between oxygen and nitrogen and should completely stop when either nitrogen or oxygen are absent. For H$_2$:He mixtures (as in the atmospheres of the planetary gas giants) a photo-ionization model has been outlined in \cite{Dubrovin:2010}, but a photo-ionization mechanism on Venus (CO$_2$:N$_2$) is unlikely to exist \cite{Dubrovin:2010}, and in pure gases it cannot exist either. An alternative is propagation through background ionization, this background can be generated by radiation or by previous discharges in a mode of repetitive pulses \cite{Pancheshnyi:2005}. Though reaction rates and electron densities ahead of the streamers can vary by several orders of magnitude between different gases, the experiments show that the propagation of positive streamers is relatively unaffected. E.g., in the recent experiments by Nijdam et al.~\cite{Nijdam:2010} the oxygen fraction changed by 5 to 6 orders of magnitude while the streamer velocity changed by less than 10\%. This suggests that positive streamers can propagate due to several mechanisms; and that they are quite robust against changes of the underlying physical mechanisms. In the present article, we test this behavior through simulations of positive streamers in different nitrogen:oxygen ratios, and through varying the parameters of photo-ionization and of background ionization.

\subsection{Photo-ionization}

Photo-ionization in air is thought to work as follows. Electrons accelerated by high electric fields in the streamer head excite electronic states of nitrogen (the species with the higher ionization energy) by impact; the excited nitrogen can then emit a photon with sufficient energy to ionize an oxygen molecule at some distance. The presence of this ionizing radiation with significant penetration length in air was demonstrated in numerous experiments \cite{Raether:1964}. A similar mechanism was proposed by Dubrovin et al.\ in section 2.2 of \cite{Dubrovin:2010} for the H$_2$:He mixtures on Jupiter-like planets; she argues that there are electronically excited He states that can emit photons with sufficient energy to ionize H$_2$ while no such mechanism exists for Venus (CO$_2$:N$_2$). However, the only quantitative photo-ionization model exists for the moment, to our best knowledge, for N$_2$:O$_2$ mixtures like air \cite{Zheleznyak:1982}. We note that this model uses excitation efficiency, quenching parameters and absorption coefficients from different experiments \cite{Teich:1967,Penney:1970} and doing so, it is not a self-consistent model. In addition, this model does not treat the appearance of various "secondary" species in the plasma region (like N and O atoms, ozone, nitrogen-oxides in the case of N$_2$-O$_2$ mixtures) which can contribute to the photo-ionization process.

In pure gases, a one-step photo-ionization scheme cannot exist. A step-wise photo-ionization process (multi-photon excitation of the same species) could be an efficient source of ionization ahead of a streamer, but such processes are much slower due to the low electron and excitation densities in streamers (see, for example, \cite{Kossyi:2005}); therefore this mechanism can not support the high propagation speed of streamer discharges.

\subsection{Detachment from background ionization}

Electron detachment from negative ions in the gas is another possible source of seed electrons. These negative (and positive) ions can appear due to various reasons. Natural radioactivity often governs the background ionization in initially non-excited gases (see \cite{Pancheshnyi:2005} and references therein). In buildings, radioactive decay of radon is the main source of ionization. The level of background ionization lies normally within $10^3$-$10^4$ positive and negative ions per cm$^3$; this is the value established by the equilibrium between ionization and recombination processes. We note that this level weakly changes with pressure, and that it can decrease due to diffusion and drift of charged species towards metal electrodes. Inside a closed metal container with controlled gas filling, the ionization density is lower.

In pulse-repetitive discharges, residual ions can be accumulated from discharge to discharge and the density of background ionization can be much higher than when it is governed by the natural sources only. A background level of about $10^7$~cm$^{-3}$ can exist for a gap of a few centimeters in air at atmospheric pressure at a 1~Hz repetition rate according to simple theoretical estimates~\cite{Pancheshnyi:2005} that will be recalled in section 2.2.

Negative ions themselves (as well as positive ions) cannot create ionization while moving in an electric field (for the range of reasonable electric fields existing at elevated pressures), while they can be a source of electrons. These free electrons appear in collisions of negative ions with other gas species. For the case of oxygen mixtures, the rate of detachment was measured as a function of applied electric field at low pressures \cite{O'Neil:1973,Goodson:1974}. Measurements at elevated pressures in air \cite{Doussot:1982} demonstrate an even higher efficiency of the detachment processes, probably due to oxygen atoms \cite{Aleksandrov:2008} and vibrationally excited species \cite{Aleksandrov:2009}. 

It must be noted that both mechanisms of electron production ahead of positive streamers exist normally even in "pure" electropositive gases. The level of impurities in the experiments \cite{Nijdam:2010} was kept below 1~ppm with much effort (carefully designed vacuum vessel, no plastic parts except for the o-ring seals, baking to reduce outgassing); and relative impurity concentrations of 10$^{-4}$ or higher are much more frequent. In both cases, these impurities include, among others, electronegative admixtures at densities sufficient for both photo-ionization and detachment to produce a sufficient level of seed electrons, since a 1~ppm level at atmospheric pressure is still $10^{13}$ particles per cm$^{3}$. Therefore it is not sufficient to simply model a pure gas without contamination.

\subsection{Goal and organization of the paper}

We investigate the role of photo-ionization versus background ionization for the propagation of positive streamers in artificial air and in nitrogen with 1 ppm oxygen through simulations. We also briefly test the case of 1 ppb oxygen in nitrogen.

In section 2, we describe the model used for our simulations. First we detail the physical model and the relevant processes and their parameters. We discuss the relation between repeated discharges and background ionization levels. Then we provide some details of the numerical implementation. Section 3 contains the results of simulations in air and their interpretation as well as a quick comparison of different photo-ionization models. Section 4 covers results in N$_2$ with small (1~ppm or less) admixtures of O$_2$. In section 5, the numerical results are compared to experiments, first by comparing streamer properties such as velocity and width, followed by a discussion on the presence of feather-like structures in streamers and their cause. Finally, we present our conclusions and an outlook on future research.

\section{Model}

\subsection{Structure of discharge model}

We simulate streamers in N$_2$:O$_2$-mixtures with mixing ratios 80:20 for artificial air and 99.9999:0.0001 for pure nitrogen with a 1 p.p.m. (a relative concentration of $10^{-6}$) contamination of oxygen. We study the role of photo-ionization and of varying levels of background ionization; negative background ions can deliver free electrons through detachment in sufficiently high electric fields. The model is a density model for the electrons, the positive ions N$_2^+$, O$_2^+$ and the negative ions $\mbox{O}_2^-$, $\mbox{O}^-$ in a given N$_2$:O$_2$ gas mixture. The space charge densities are coupled to the electric field, and the reactions are specified in the next subsection. The ionization density stays so low that the change of neutral particle densities can be neglected. All ions are approximated as immobile on the time scale of the simulation; therefore their densities change only due to reactions. Electrons drift in the electric field and diffuse. Therefore the model is written as
\begin{eqnarray}
\frac{\partial n_e}{\partial t} & = & \nabla \cdot (n_e \mu_e \mathbf{E}) + D_e \nabla^2  n_e + S_e,\\
\frac{\partial n_i}{\partial t} & = & S_i,\qquad i=1,\ldots,N,
\end{eqnarray}
where $n_e$ and $n_i$ are the local number densities of the electrons or of the $N$ ion species labeled by $i$. $\mu_e$ and $D_e$ are the electron mobility and diffusion coefficients taken from \cite{Davies:1971} and \cite{Dutton:1975}, respectively.  $\mathbf{E}$ is the local electric field. The source terms $S_e$ or $S_i$ contain all production or loss reactions for the electrons or the ions of species $i$.
The local space charge density $q$ is the sum of the individual charge densities of all particles,
\begin{equation} q = \displaystyle\sum_i q_i n_i - e n_e, \end{equation}
where $q_i = \pm e$ is the charge of ion species $i$, and $e$ is the elementary charge. The electric field is coupled to the charge density through the Poisson equation
\begin{equation}\epsilon_0 \nabla \cdot \nabla \phi = - q;\end{equation}
We calculate in electrostatic approximation
\begin{equation}{\bf E}=-\nabla \phi.\end{equation}

\subsection{Modeling the reactions, including electron detachment and photo-ionization}

\subsubsection{Reactions}

The reactions included in the model are impact ionization of nitrogen and oxygen, photo-ionization, attachment of electrons to oxygen, detachment of electrons from $\mbox{O}_2^-$, electron-ion- and ion-ion-recombination; they are listed in Table~\ref{tab:reactions}. The reaction rates depend on the densities of the interacting species and a field-dependent rate coefficient. The rate coefficients for impact ionization, electron attachment and recombination are based on the kinetic model of Kossyi \textit{et al.}~\cite{Kossyi:1992} with some of the rate coefficients generated by the BOLSIG+ Boltzmann-solver \cite{Hagelaar:2005}. The electron detachment rates are taken from Kossyi \textit{et al.}~\cite{Kossyi:1992} and are discussed in more detail by Capitelli \textit{et al.}~\cite{Capitelli:2000}. The photo-ionization model is from Luque \textit{et al.}~\cite{Luque:2007}. Both detachment and photo-ionization can be a source of free electrons ahead of the streamer; therefore we discuss them now in more detail.

\begin{table}
\begin{tabular}{| l | l |}
\hline
\textbf{Reactions} & \textbf{Reference}\\
\hline
Impact ionization: & \\
$\mbox{e}^- + \mbox{N}_2 \rightarrow \mbox{e}^- + \mbox{e}^- + \mbox{N}_2^+$ & BOLSIG+ \cite{Hagelaar:2005}\\
$\mbox{e}^- + \mbox{O}_2 \rightarrow \mbox{e}^- + \mbox{e}^- + \mbox{O}_2^+$ & BOLSIG+\\
\hline
Photo-ionization: & \\
$\mbox{e}^-+\mbox{N}_2 \to \mbox{e}^-+\mbox{N}_2^*+\mbox{UV-photon},$ & Luque {\it et al.} \cite{Luque:2007}\\
$\qquad \mbox{then UV-photon}+\mbox{O}_2 \to \mbox{O}_2^++\mbox{e}^-$ & \\
\hline
Attachment of electrons: & \\
$\mbox{e}^- + \mbox{O}_2 + \mbox{O}_2 \rightarrow \mbox{O}_2^- + \mbox{O}_2$ & BOLSIG+\\
$\mbox{e}^- + \mbox{O}_2 \rightarrow \mbox{O} + \mbox{O}^-$ & BOLSIG+\\
$\mbox{e}^- + \mbox{O}_2 + \mbox{N}_2 \rightarrow \mbox{O}_2^- + \mbox{N}_2$ & Kossyi \textit{et al.}
\cite{Kossyi:1992}\\
\hline
Detachment of electrons: & \\
$\mbox{O}_2^- + \mbox{O}_2 \rightarrow \mbox{e}^- + \mbox{O}_2 + \mbox{O}_2$ & Capitelli \textit{et al.} \cite{Capitelli:2000}\\
$\mbox{O}_2^- + \mbox{N}_2 \rightarrow \mbox{e}^- + \mbox{O}_2 + \mbox{N}_2$ & Capitelli \textit{et al.}\\
\hline
Recombination: & \\
$\mbox{e}^- + \mbox{X}^+ \rightarrow \mbox{neutrals}$ & Kossyi \textit{et al.}\\
$\mbox{O}^- + \mbox{X}^+ \rightarrow \mbox{neutrals}$ & Kossyi \textit{et al.}\\
$\mbox{O}_2^- + \mbox{X}^+ \rightarrow \mbox{neutrals}$ & Kossyi \textit{et al.}\\
$\mbox{O}^- + \mbox{X}^+ + \mbox{X} \rightarrow \mbox{neutrals}$ & Kossyi \textit{et al.}\\
$\mbox{O}_2^- + \mbox{X}^+ + \mbox{X} \rightarrow \mbox{neutrals}$ & Kossyi \textit{et al.}\\
\hline
\end{tabular}
\caption{Overview of the reactions included in our model and the references for their rate coefficients. X denotes any neutral species and consequently, X$^+$ denotes any positive ion.}\label{tab:reactions}
\end{table}

\subsubsection{Detachment}

\begin{figure}
\center{\includegraphics[width=0.5\columnwidth]{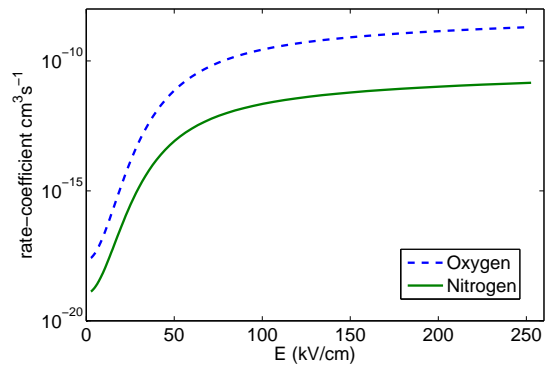}}
\caption{Rate coefficients for detachment mediated by two different neutral species as a function of the local electric field. The dashed curve shows the rate coefficient for detachment via collision with O$_2$ (8), the solid curve shows the rate coefficient for detachment via collission with N$_2$ (9). All data is for standard temperature and pressure.}\label{fig:detach}
\end{figure}

Electron detachment from O$_2^-$ can occur when the negative ion collides with a neutral gas particle. The rate at which electrons detach from negative ions depends on the collision frequency, on the local electric field and on the density of the neutral gas. Since we consider N$_2$:O$_2$ mixtures, we have two separate detachment reactions as listed in table~\ref{tab:reactions}. The rate coefficients for these reactions are plotted in figure~\ref{fig:detach}.

Various processes can contribute to the presence of O$_2^-$ in the gas. First of all, external sources such as radioactive materials in buildings (primarily radon) and cosmic rays can generate an equilibrium level of background ionization. This level is estimated to be 10$^3$-10$^4$~cm$^{-3}$ \cite{Pancheshnyi:2005}. A second source of background ionization is the residual ionization from a previous discharge. In many experiments and practical applications, discharges are generated in a repetitive way. By the time the next discharge is started, some residual ionization still remains in the form of O$_2^-$ and positive ions.

We here give a simple estimate for the level of background ionization in the region far from the needle electrode in a repetitive discharge with repetition frequency in the range of 1 Hz. The streamers then explore a different part of space in each consecutive voltage pulse \cite{Nijdam:2010}. If the electron density is $10^{14}$~cm$^{-3}$ in the streamer and a fraction of $10^{-3}$ of space is filled with streamer plasma, then the average electron density in space is $10^{11}$~cm$^{-3}$. Now these electrons will attach rapidly to O$_2$ to form O$_2^-$, even with 1~ppm of O$_2$, the attachment time is only 20~ms \cite{Pancheshnyi:2005}. The density of both negative and positive ions will then be $10^{11}$~cm$^{-3}$ after diffusion has smeared out the original streamer structure. The bulk recombination rate $\beta$ is approximately 10$^{-6}$-10$^{-7}$~cm$^3$/s \cite{Kossyi:1992,Capitelli:2000}. We then arrive at the following differential equation for the ion densities ($n_+ = n_- = n$):
\begin{equation}\partial_t n = -\beta n^2,\end{equation}
which is solved by
\begin{equation}n(t) = \frac{1}{\frac{1}{n(0)} + \beta t} \simeq \frac{1}{\beta t}\ \textrm{(for }t > 1\textrm{~ms)}.\end{equation}
Therefore one expects the level of O$_2^-$ to be around 10$^6$-10$^7$~cm$^{-3}$ when the repetition frequency of the discharges is 1~Hz.

The precise detachment rate \cite{Aleksandrov:2008,Aleksandrov:2009} as well as the evolution of negative ions in the plasma in the time between two streamer discharges \cite{Popov:2010} is presently under debate and future simulations might benefit from the inclusion of a more detailed chemistry model. In our model, the critical electric field at which the rate coefficient for detachment equals the rate coefficient for attachment is 70~kV/cm in air. Since the attachment rate is orders of magnitude lower in nitrogen with 1~ppm oxygen and the detachment rate is not as strongly dependent on the electric field as, for example, the impact ionization rate, the critical field for detachment in nitrogen with 1~ppm oxygen is only 20~kV/cm.

\subsubsection{Photo-ionization}

Photo-ionization in air is based on the fact that there are excited states of nitrogen molecules that can relax through emission of a UV-photon with energy high enough to ionize an oxygen molecule. The history of the concept and the present (poor) data situation was recently reviewed in~\cite{Nijdam:2010}. Typically, photons in two or three spectral ranges are included; the most used model is presently the one of Zheleznyak \textit{et al.}~\cite{Zheleznyak:1982}. Zheleznyak \textit{et al.} merged the available experimental data to create this model for the process:
\begin{equation}
S_{ph}(\mathbf{r}) = \frac{\xi}{4\pi} \frac{p_q}{p + p_q} \int \frac{h(p |\mathbf{r} - \mathbf{r'}|) S_i(\mathbf{r'}) d^3 (p\mathbf{r'})}{|p \mathbf{r} - p \mathbf{r'}|^2},\label{eqn:photo_integral}
\end{equation}
where $\xi$ is a proportionality constant, $p$ is the gas pressure, $p_q = 80$~mbar the quenching pressure, $S_i$ is the local impact ionization rate of nitrogen and $h$ the absorption function of the ionizing photons. Since integral expressions such as these are computationally costly to solve, Luque \textit{et al.} approximated equation~(\ref{eqn:photo_integral}) by a set of Helmholtz differential equations \cite{Luque:2007} (and in parallel Bourdon \textit{et al.}~\cite{Bourdon:2007} did the same):
\begin{equation}
S_{ph} = \frac{p_q}{p + p_q} \displaystyle \sum_{j = 1}^N A_j S_{ph,j},\hspace{3mm} (\nabla^2 - \lambda_j^2) S_{ph,j} = S_i
\end{equation}
where $A_j$ and $\lambda_j$ are chosen to fit the experimental model as well as possible. $\lambda_j$ is related to the characteristic absorption length and $A_j$ represents an intensity. Unless otherwise specified in the paper, we have used the original fit by Luque \textit{et al} with two Helmholtz terms and the following parameters: $A_1$ = $4.6 \times 10^{-2}$~cm$^{-1}$~bar$^{-1}$, $A_2$ = $2.7 \times 10^{-3}$~cm$^{-1}$~bar$^{-1}$, $\lambda_1 = 45$~cm$^{-1}$~bar$^{-1}$ and $\lambda_2 = 7.6$~cm$^{-1}$~bar$^{-1}$. (The quantities were originally given in Torr$^{-1}$ rather than in bar$^{-1}$.) All our simulations were conducted at standard temperature and pressure.

The longest of the absorption lengths is the main contribution to the non-local effects of the photo-ionization. In the fit by Luque \textit{et al.}~\cite{Luque:2007}, the longest of the two absorption lengths is 1.3~mm in air at standard temperature and pressure. The absorption length scales inversely with the density of oxygen molecules in the mixture: in nitrogen with 1~ppm O$_2$, the oxygen density is $2 \times 10^5$ times lower than in air and consequently, the absorption length is $2 \times 10^5$ times longer: 260~m. We note that the Zheleznyak model and the approximations that are based on it assume that the UV-photons are emitted instantenously when the molecule is excited. This assumption is implemented in all streamer simulations we know of and its results are in good agreement with experiments. It been remarked in \cite{Steine:1999, Liu:2004} and by C. Li (unpublished) that the non-instantenous emission can cause some retardation in the photo-ionization process. However, estimates for retardation times are not well based yet. A full model of the population dynamics of the excited states would be required to accurately predict this retardation.

\subsection{Electrode geometry, voltage and initial conditions} \label{sec:initial}

\begin{figure}
\begin{center}
\includegraphics[height=7cm]{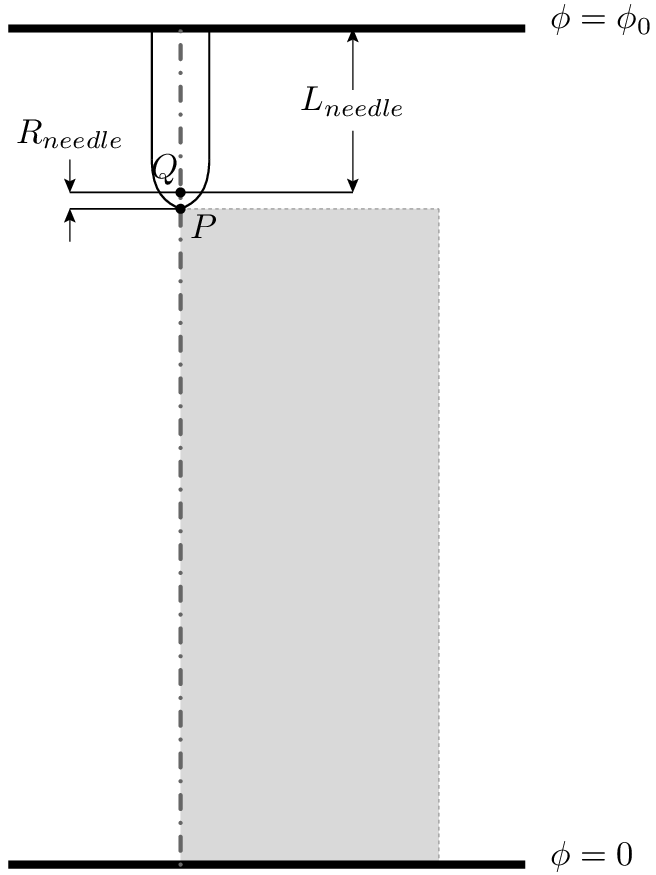}
\caption{Schematic of the computational setup. The shaded rectangle represents the computational domain for the fluid equations, the thick horizontal lines the 2 planar electrodes with the needle and its parameters depicted at the anode. The area between the two planar electrodes is the computational domain for the Poisson equation. The needle is simulated by a single point charge, $Q$, chosen such that $\phi = \phi_0$ in the point $P$, which is the tip of the needle. The calculation assumes cylindrical symmetry around the needle axis represented by the dashed-dotted line.}\label{fig:setup}
\end{center}
\end{figure}

We simulate a needle-plane electrode configuration as shown in figure~\ref{fig:setup}. The needle electrode protrudes from a planar electrode and is positively charged while the planar electrode below is grounded, resulting in an electric field (in the absence of space charges) pointing from the needle towards the plane. The voltage is constant throughout the simulation. The gap between the tip of the electrode needle and the planar electrode is 4~mm or 8~mm. The length of the needle $L_{needle}$ is 2~mm and its radius $R_{needle}$ is 0.2~mm. The potential between the two electrodes is fixed to 12~kV for the 4~mm gap and to 20~kV for the 8~mm gap. The radius of the computational domain is 2~mm (4~mm gap) or 3~mm (8~mm gap).

As an initial condition, we place an electrically neutral seed of electrons and positive ions at the tip of the needle. The seed consists of half of a Gaussian with the peak located at the needle tip. The maximal density of this initial seed is $3.4 \times 10^7$~cm$^{-3}$. The width of the Gaussian (the distance at which the density fallen to a factor of $1/e$ of the maximum) is $73.6$ $\mu$m. As was shown by Luque \textit{et al.} in \cite{Luque:2008}, the density of the initial seed hardly influences a positive streamer when it starts from a pointed electrode. In the cases where field detachment is studied, a uniform and electrically neutral density of $\mbox{O}_2^-$ ions and positive ions is added.

\subsection{Numerical implementation}

We assume cylindrical symmetry of the simulated system. As a consequence, only the radial and longitudinal coordinates $r$ and $z$ are considered. We use the numerical code developed by Montijn \textit{et al.}~\cite{Montijn:2006} and extended with photo-ionization by Luque \textit{et al.}~\cite{Luque:2007}; it uses an adaptive grid-refinement scheme to increase the spatial resolution where necessary: most notably in the head of the streamer. Different grids with different refined areas are used for the particle densities and for the electric field.

The needle electrode is modeled by a floating point charge using a ``charge simulation technique'' as described in \cite{Luque:2008}, and earlier in \cite{Singer:1974}. The computational domain of the density equations starts at the tip of the needle electrode and extends towards the planar electrode, depicted by the shaded area in figure~\ref{fig:setup}. The computational domain for the Poisson equation is the region between the two planar electrodes, including the simulated needle. This area is depicted in figure~\ref{fig:setup} by the area between the two bold horizontal lines and to the right of the vertical dashed-dotted line.

The computational domain for the density equations is initially covered by a rectangular grid of 360 $\times$ 200 cells for the 4~mm gap and 720 $\times$ 300 cells for the 8~mm gap, resulting in a cell-size of approximately 11.1~$\mu$m $\times$ 10~$\mu$m at the coarsest level. At every level of refinement, the refined grid contains cells of half the length and height of the cells at the coarser level. We have used up to 3 levels of refinement, leading to a cell-size at the finest level of approximately 1.39~$\mu$m $\times$ 1.25~$\mu$m. The criteria for refinement are such that the head of the streamer is always in the area of maximal refinement, though this area is not necessarily restricted to the streamer head. In figure~\ref{fig:setup}, the computational domain of the density equations is depicted by the gray area.

For the density equations we use homogeneous Neumann boundary conditions at the top, bottom and outer edges of the domain. A homogeneous Neumann boundary condition represents the symmetry on the central axis. For the Poisson equation as well as for the Helmholtz equations calculating the photo-ionization, we use homogeneous Dirichlet boundary conditions at the top, bottom and outer edges of the domain and again a symmetric Neumann boundary condition on the central axis. The homogeneous electric field created by the planar electrodes is added in a second step. Note that the top boundary for the Poisson equation is not the same as the top boundary for the density equations. 

\section{Simulations in air: photo-ionization versus background ionization}

\subsection{Either photo-ionization or background ionization}

We here consider streamers in artificial air which is a mixture of 80\% $\mbox{N}_2$ molecules and 20\% $\mbox{O}_2$ molecules. We use standard temperature and pressure (STP), i.e., the pressure is 1 bar and the temperature 300~K. The distance between the two planar electrodes is 6~mm with a 2~mm needle protruding from the anode. Consequently, the propagation length of the streamer and the length of the computational domain for the density equations is 4~mm. The applied voltage is 12~kV, therefore the average field between the planar electrodes is 20~kV/cm. The initial electron and ion density near the electrode needle is described in subsection~\ref{sec:initial}. We consider four scenarios:
\begin{enumerate}[1.]
\item Photo-ionization, no initial background ionization.
\item No photo-ionization, initial uniform background ionization $[\mbox{O}_2^-] = 10^7$ cm$^{-3}$.
\item No photo-ionization, initial uniform background ionization $[\mbox{O}_2^-] = 10^5$ cm$^{-3}$.
\item No photo-ionization, initial uniform background ionization $[\mbox{O}_2^-] = 10^3$ cm$^{-3}$.
\end{enumerate}
The first scenario is the usual streamer model in air with the standard photo-ionization model for N$_2$:O$_2$ mixtures. In the other scenarios, photo-ionization is excluded, but different levels of background ionization are included. The second scenario corresponds with a streamer in a series of repeated discharges of approximately 1~Hz while the fourth scenario represents the background ionization present due to ambient sources such as radioactive materials in buildings, see discussion in section 2.2.

\begin{figure}
\center{\includegraphics[width=0.5\columnwidth]{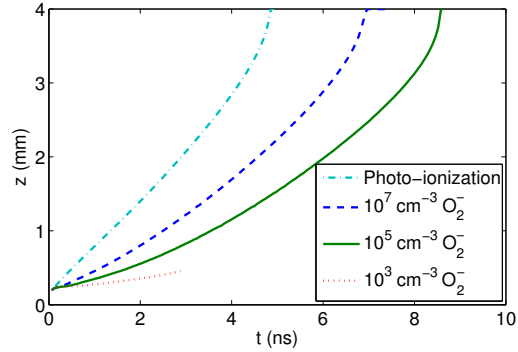}}
\caption{Position of the streamer head as function of time. The top curve corresponds to an air streamer with photo-ionization (scenario 1), the curves below that to streamers without photo-ionization, and with a decreasing amount of background ionization (scenarios 2, 3 and 4). The $z = 0$ point corresponds to the tip of the needle.}\label{fig:air_pos_ov}
\end{figure}

Figure~\ref{fig:air_pos_ov} shows the position of the streamer head as a function of time for the 4 scenarios. The position of the streamer head is defined as the position of the charge maximum on the axis of symmetry. At the start of the simulation, both the positively charged and the negatively charged initial seeds are equal and no space charge is present. Immediately thereafter, however, charges separate under the influence of the electric field and a space charge layer is formed at some distance from the origin. This explains the initial jump in position that can be seen in figure~\ref{fig:air_pos_ov} as well as the increased initial velocity in later figures.

We see that streamers with photo-ionization are the fastest under the conditions of the present simulations. But a sufficiently high level of background ionization, e.g., in a repetitive discharge produces enough free electrons in front of the streamer head for the streamer to propagate. The streamer propagates about 40\%  faster with photo-ionization than with a background ionization of $10^7$ cm$^{-3}$; according to the discussion in section 2.2, this is roughly the background ionization for a discharge with repetition frequency of 1~Hz. When the background ionization density decreases by a factor of 100 from 10$^7$ to 10$^5$~cm$^{-3}$ (corresponding to a frequency of 0.01 Hz), the time it takes for the streamer to cross the gap increases by only $20\%$. However, a background ionization density of $10^3$ cm$^{-3}$ was not sufficient to start a streamer that would propagate more than a few hundreds of micrometers under the modeled circumstances (the width of the initial Gaussian seed is 70~$\mu$m); this background density characterizes ambient air at ground level without previous discharges. Note that as the streamer approaches the cathode, the electric field in front of the streamerhead increases, since the total potential remains unchanged and the streamer interior is almost completely screened. As a consequence, streamers accelerate when they approach the cathode.

\begin{figure}
\center{\includegraphics[width=0.5\columnwidth]{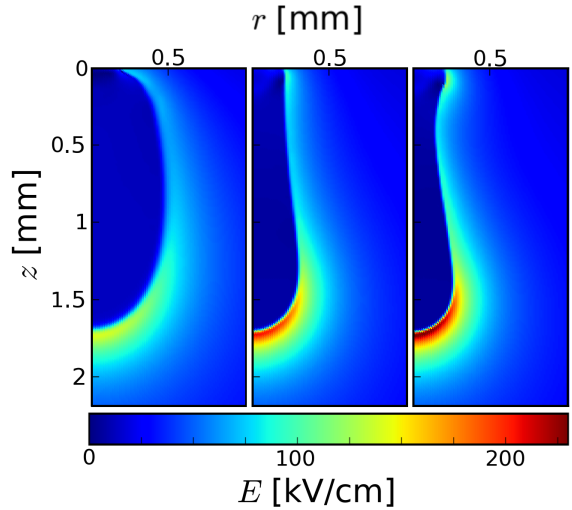}}
\caption{Absolute value of the electric field for streamers with photo-ionization (left, scenario 1), 10$^7$ cm$^{-3}$ background ionization (middle, scenario 2) and 10$^5$ cm$^{-3}$ background ionization (right, scenario 3). The images are taken after simulation times of approximately 3, 4 and 6~ns, respectively. The computational domain is larger than the plotted area.}
\label{fig:air_crosssect}
\end{figure}

Streamer diameter and shape and maximum value of the enhanced electric field are shown in Figure~\ref{fig:air_crosssect}. When photo-ionization is present, the streamer is wider by a factor of 2 and the maximal electric field is 30\% lower than in the case of background ionization. In addition, the electric field profile is less steep with photo-ionization than with background ionization. Streamers with photo-ionization are wider and smoother than those with background ionization, because detachment from background ionization is mostly a local effect (the critical electric field for detachment in air is 70~kV/cm in our model), while photo-ionization is highly non-local with a characteristic absorption length of the ionizing photons of 1.3~mm in air at atmospheric pressure (cf. section 2.2). Therefore, free electrons are generated in a much larger region by photo-ionization than by detachment; this will be illustrated later in Fig.~\ref{fig:n2p_n2i_airp_elec_dens}.

It is worth noting that although the background ionization differs by 2 orders of magnitude between scenarios 2 and 3, the diameter of the streamer and the maximal electric field are practically the same when the streamer head has reached the same point, though the evolution times differ. However, a certain minimum level of background ionization is required to generate propagating streamers. For this reason, scenario 4 was omitted from Figure~\ref{fig:air_crosssect}, because the streamer did not propagate sufficiently far, in accordance with Fig.~\ref{fig:air_pos_ov}.

\subsection{Combining photo-ionization and background ionization}

To determine the relative influence of photo-ionization and background ionization, we run another simulation that combined scenarios 1 and 2: a $10^7$ cm$^{-3}$ density of $\mbox{O}_2^-$ was added to a model with photo-ionization. The results of this simulation were virtually indistinguishable from the results of scenario 1 with photo-ionization only. We therefore conclude that since both mechanisms are present in air, photo-ionization dominates over the effect of detachment from background ionization for the generally accepted photo-ionization model.

\begin{figure}
\center{\includegraphics[width=0.5\columnwidth]{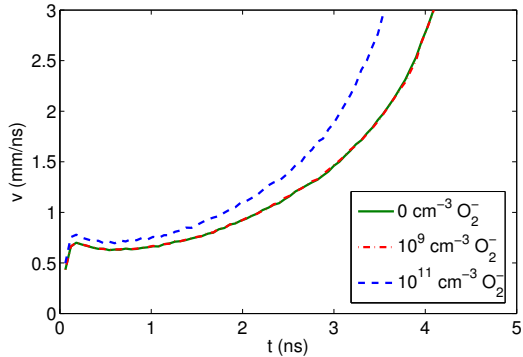}}
\caption{Propagation speed of streamers in air with different levels of background ionization. Photo-ionization is present in all 3 cases.}
\label{fig:air_photo_ionlevels}
\end{figure}

As discussed in section 2.2, in discharges with repetition frequencies as high as 1~kHz the background ionization can reach a level of 10$^{10}$~cm$^{-3}$. We therefore have investigated two additional cases with photo-ionization as well as background ionization. Figure~\ref{fig:air_photo_ionlevels} shows, that only at a level of 10$^{11}$~cm$^{-3}$ negative oxygen ions starts to increase the streamer velocity. At lower levels of O$_2^-$, streamers propagate due to photo-ionization and are insensitive to the additional background ionization. We remark that at these high repetition frequencies, there may not have been enough time for the residual ionization to diffuse into a homogeneous density distribution. This may cause memory-effects, where streamer propagation is easier over a path taken by the previous discharge.

\subsection{Testing different photo-ionization models}

\begin{figure}
\center{\includegraphics[width=0.5\columnwidth]{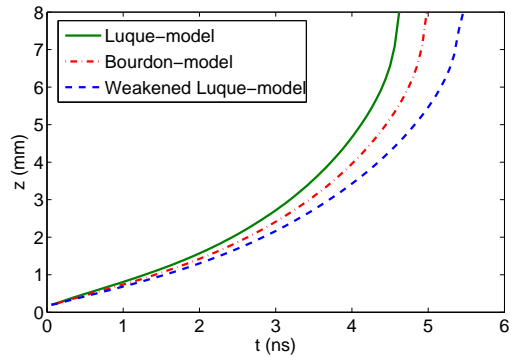}}
\caption{Comparison of three photo-ionization models. The curve labeled ``Luque-model'' refers to scenario 1 and uses the photo-ionization model by Luque \etal \cite{Luque:2007}. The curve labeled ``Bourdon-model'' shows the result of a 3-term Helmholtz model for photo-ionization by Bourdon \etal \cite{Bourdon:2007} as opposed to our default 2-term model. The bottom curve uses the Luque-model, where the source term for the photo-ionization equation was artificially reduced by a factor of $10$. Note that the weakened scenario does not represent an actual physical scenario; its purpose is to demonstrate the influence of the accuracy of the photo-ionization parameters. }\label{fig:air_photo_comp}
\end{figure}

It is well known~\cite{Pancheshnyi:2005,Nijdam:2010,Nudnova:2008}, that the actual parameters of photo-ionization are not very well known. We therefore test here how much the simulation results depend on the parameters of the photo-ionization model. We compare streamers in air without background ionization in three cases. The first includes photo-ionization according to Luque's approximation~\cite{Luque:2007}. Bourdon {\it et al.}~\cite{Bourdon:2007} suggest that the 2-term Helmholtz model for photo-ionization is insufficient and propose to replace it by a similar model with 3 terms. This 3-term model we call the Bourdon model here. The final test-case uses our default 2-term model, but with the number of emitted photons artificially reduced by a factor of 10; this serves as a model for our lack of knowledge of the actual parameters. The longest characteristic absorption length in Bourdon's 3-term model is 0.5~mm in STP air, while it is 1.3~mm in Luque's 2-term model.

The gap between the electrodes is here increased to 8~mm so that any differences would have time to develop; and the applied voltage is now 20~kV. The positions of the streamer heads as a function of time are plotted in figure~\ref{fig:air_photo_comp}; it shows that the difference between the unaltered scenario (which is identical to the aforementioned scenario 1, only with a larger electrode gap) and the weakened one is rather small: The weakened scenario has $10$ times less source electrons in front of the streamer head, but it only takes $20\%$ longer to cross the gap between the electrodes. The results with Bourdon's 3-term approximation lie between the other two curves, i.e., they deviate from the results with Luque's photo-ionization model by 10\%.

\subsection{Summary of results in air}

With our simulations in air, we have found that while background ionization levels of 10$^5$~cm$^{-3}$ or more can provide sufficiently many electrons for positive streamers to propagate in the absence of photo-ionization, in real-life experiments and applications, the photo-ionization mechanism will dominate the effects of any background ionization level of 10$^9$~cm$^{-3}$ or less. Therefore, we only expect to see the effects of background ionization on streamer propagation in experiments with repeated discharges with a repetition frequency of 1~kHz or more.

Additionally, we found that positive streamers in air are remarkably insensitive to the precise conditions of the source of the seed electrons (both the mechanism and the number of electrons produced). Changing the background ionization level by two orders of magnitude only resulted in a 20\% difference in the time it takes to cross the electrode gap. Also, changing the photo-ionization model to an artificial one with ten times less photo-ionization events has similarly small effects on streamer propagation.

\section{Simulations in ``pure'' $\mbox{N}_2$}

We now investigate streamers in nitrogen of high purity. In Nijdam's experiments \cite{Nijdam:2010}, the impurity level was kept below 1 ppm. We here simulate an admixture of 1 ppm oxygen in nitrogen. The different ratio of nitrogen and oxygen changes both the number of emitted photons and their absorption lengths.
The artificially weakened photo-ionization from section 3.3 amounts to lowering the $\mbox{N}_2$ density for the purpose of photo-ionization, while keeping the overall gas pressure constant, as it results in a lower number of emitted photons. The other parameters in our photo-ionization model are the absorption lengths of the ionizing photons of different wave lengths. These lengths are inversely proportional to the $\mbox{O}_2$ density. The longest of these, denoted by $l_a$, has the strongest effect on the non-local characteristics of the process. In artificial air at atmospheric pressure, we have $l_a = 1.3$~mm. In $\mbox{N}_2$ with a 1~ppm admixture of $\mbox{O}_2$, this absorption length is increased to 260~m. In cases where $l_a$ is much larger than the size of the modeled domain, the decay profile of $\mbox{O}_2$ ionization events as function of distance from the photon source (the streamer head) is dominated by a $r^{-2}$ falloff, with $r$ the distance to the photon source.

\begin{figure}
\center{\includegraphics[width=0.5\columnwidth]{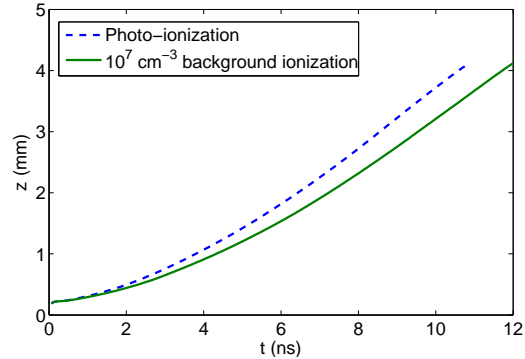}}
\caption{Position of streamer head as a function of time for two cases in $\mbox{N}_2$ with 1ppm $\mbox{O}_2$. Solid curve: $10^7$ cm$^{-3}$ background ionization, no photo-ionization. Dashed curve: photo-ionization, no background ionization. The gap between the electrodes is 8~mm and the potential is 20~kV. For both cases, the available data was limited by the fact that branching occurs at around 4~mm and the simulations were halted before the streamers reached the cathode.}\label{fig:n2_pos_overview}
\end{figure}

Figure~\ref{fig:n2_pos_overview} shows that in $\mbox{N}_2$ with a 1 ppm admixture of $\mbox{O}_2$, both photo-ionization as well as a background ionization of 10$^7$ cm$^{-3}$ can produce streamers. The propagation speeds are more similar than they are for the same scenarios in artificial air, which is an indicator of the lowered effect of photo-ionization: compared to air, the amount of ionizing photons produced is $25\%$ higher in $\mbox{N}_2$, but the characteristic absorption length is $2 \times 10^5$ longer.

\begin{figure}
\center{\includegraphics[width=0.5\columnwidth]{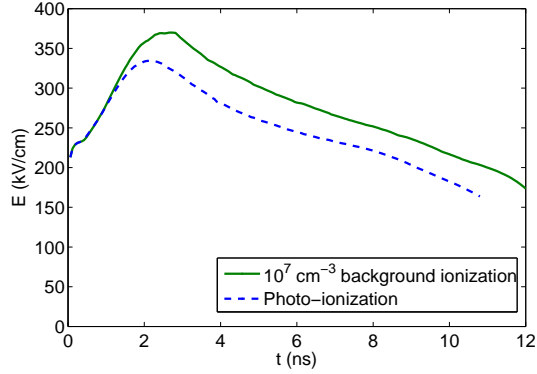}}
\caption{Maximal electric field on the symmetry axis of the streamer as a function of time for two cases in $\mbox{N}_2$ with 1ppm $\mbox{O}_2$. Solid curve: $10^7$~cm$^{-3}$ background ionization, no photo-ionization. Dashed curve: photo-ionization, no background ionization. The gap between the electrodes is 8 mm.}\label{fig:n2_eabs}
\end{figure}

Just like the propagation speed, the maximal of the electric field and its development over time is very similar for both two mechanisms, as can be seen in figure~\ref{fig:n2_eabs}. Background ionization gives rise to slightly higher (between 15\% and 20\%) fields, but the evolution of the field strength in time remains the same: The field starts at a base value determined by electrode and its applied voltage, then it rises to a maximum and drops as the streamer becomes less focused until it finally branches. The branching sets in after the streamer has propagated 3.7~mm (photo-ionization case) or 4.2~mm (background ionization case). At the onset of the branching event, the rounded space charge layer becomes increasingly flat, leading to a more strongly enhanced electric field at the corners and consequently propagation in a direction that deviates from the axis, this was seen similarly in \cite{Ebert:2006, Luque:2007, Kao:2010}. Simulations are halted once branching occurs, as this breaks the cylindrical symmetry of the system.

As the source of the background ionization level of 10$^7$~cm$^{-3}$ is the residual ionization from repeated discharges, we expect that in virgin air the streamer would only propagate due to the photo-ionization mechanism. In simulations where photo-ionization and background ionization were combined, the presence of background ionization on the streamer velocity was noticeable at levels of 10$^7$~cm$^{-3}$ and higher. From this we can conclude that in nitrogen with 1~ppm oxygen, the effect of the repetition frequency can be seen with repetition frequencies of 1~Hz or higher, as our estimate for ion densities (cf. section 2.2) applies equally in air and in N$_2$ with 1~ppm O$_2$.

\begin{figure}
\center{\includegraphics[width=0.5\columnwidth]{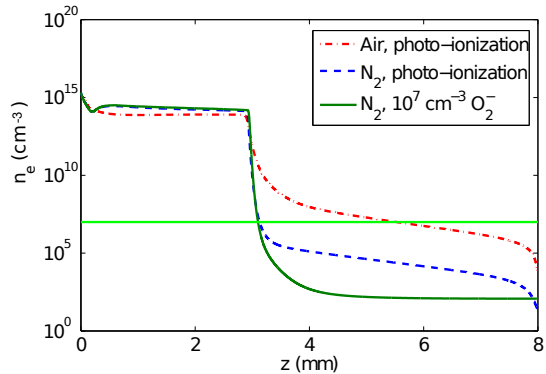}}
\caption{Electron density on the streamer axis where the streamer propagates from left to right. The solid curve shows $\mbox{N}_2$ with 1~ppm O$_2$ with a background ionization of $10^7$ cm$^{-3}$, the dashed curve shows $\mbox{N}_2$ with 1~ppm O$_2$ with photo-ionization and the dashed-dotted curve shows air with photo-ionization. All three curves represent a streamer of equal length and are therefore from different time steps. Also included is the initial level of O$_2^-$ background ionization (horizontal line).}\label{fig:n2p_n2i_airp_elec_dens}
\end{figure}

The non-local effect of photo-ionization is visible in the electron density profile in Figure~\ref{fig:n2p_n2i_airp_elec_dens}. While in the background ionization scenario, free electrons are only created near the streamer head, in the area with a high electric field (where the field exceeds 20 kV/cm in nitrogen with 1 ppm oxygen), the long characteristic absorption length $l_a$ of photo-ionization causes electrons to be freed at a significant distance from the streamer head. The relatively constant, non-zero electron density in the background ionization case is due to an equilibrium between detachment (rate-coefficient is low in the low-field region ahead of the streamer, but the $\mbox{O}_2^-$ density is rather high) and attachment (high rate-coefficient, but low level of $e^-$ density) of electrons.

The effect of the absorption length of the ionizing photons can be seen when comparing the curves from air and $\mbox{N}_2$ in figure~\ref{fig:n2p_n2i_airp_elec_dens}. Ahead of the ionization front, the electron density in air is 2 to 3 orders of magnitude larger than in $\mbox{N}_2$. The long absorption length of the ionizing photons in $\mbox{N}_2$ with 1 ppm $\mbox{O}_2$ (260 m) means that most of the photons do not ionize an $\mbox{O}_2$ molecule before leaving the computational domain, while in air (1.3 mm absorption length), the opposite is true.

In general, the number of photons that reaches a point at a distance $r$ from the streamer head is proportional to $S_{ph} \times e^{-r/l_a} \times r^{-n}$ where $l_a$ is the absorption length and $n \geq 0$ describes the algebraic falloff of the photon intensity. $n$ depends on the shape of the photon source. For a point-source $n = 2$, for a planar front $n = 0$. When the distance $r$ is large, the source of the photons can be approximated by a point source and we get $n = 2$. For small $r$, the complex structure of the streamer head will give rise to a smaller value of $n$. In nitrogen with a 1~ppm admixture of oxygen, the absorption length is so large compared to the size of the domain that the exponential falloff can be neglected. In air, the absorption length is still fairly large compared to the size of the domain, but its contribution can no longer be neglected. This can be seen in figure~\ref{fig:n2p_n2i_airp_elec_dens}, where the slope of the electron density curve in air is steeper than in nitrogen in the area in front of the streamer head.

With a naive comparison of the absorption lengths, one expects to see a $2 \times 10^5$ times lower electron density in nitrogen with 1~ppm oxygen than in air due to the difference in photo-ionization intensity: The number of photo-ionization events in a small test-volume is proportional to the number of photons entering this test volume multiplied by the number of oxygen molecules in the test volume. Assuming identical photon-sources and no loss of photons between the source and the test-volume, the number of photons entering the test-volume is independent of gas composition and the number of photo-ionization events is proportional to the oxygen density.

However, from figure~\ref{fig:n2p_n2i_airp_elec_dens}, this difference seems to be about $10^3$. There are several reasons that explain the difference between the naive expectation and the obtained results. First, photon absorption in air diminishes the photon number due to the non-negligible exponential falloff $e^{-r/l_a}$ as described in the previous paragraph. Second, the field at the streamer head is higher in nitrogen, which, along with the 25\% higher N$_2$ density causes a higher number of photons to be emitted. And finally, in air a large part of the photo-electrons is lost due to attachment to oxygen, whereas in nitrogen this attachment is orders of magnitude lower due to the lower oxygen density.

The sudden drop of the electron density near the cathode is due to the choice of a Dirichlet boundary condition for the photon density and the fact that reactions and ionization on and in the electrode are not modeled. The production of electrons due to photo-ionization falls off evenly, but the background electric field moves the electrons towards the anode. Everywhere else in the domain, the balance between electrons moving away from a point and those moving towards it results in a relatively flat density profile, but near the cathode, at any point, electrons that move away from a point are not replaced by electrons coming from the cathode. In reality, one might expect an increase in electron density near the electrode due to photo-emissions of electrons from the metal in the electrode, but this goes beyond the scope of this paper.

\begin{figure}
\center{\includegraphics[width=0.5\columnwidth]{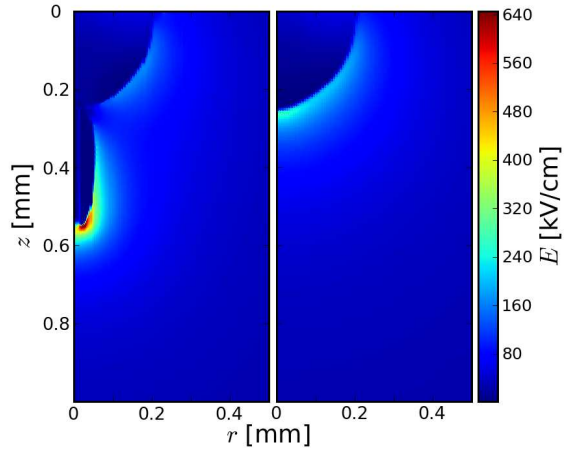}}
\caption{Electric field strength of simulated streamers in $\mbox{N}_2$ with 1ppb $\mbox{O}_2$ contamination. Both cases have photo-ionization, but no background ionization. The physical parameters of the two cases are identical with the exception of the size of the discharge gap. Compared to the left picture, the right picture has a shorter gap (still many times larger than the scale at which space charge processes occur), and therefore a smaller computational domain, but twice the spatial resolution on the coarsest level (5.56~$\mu$m versus 11.1~$\mu$m) and four times the spatial resolution on the finest level (0.347~$\mu$m versus 1.39~$\mu$m). We conclude that the propagating streamer in the left panel is due to a numerical artifact. (In both cases, the computational domain is larger than the plotted area.)}\label{fig:n2_1ppb}
\end{figure}

To investigate how much photo-ionization was actually required to yield a propagating streamer, we decreased the level of oxygen contamination in the simulation. As the oxygen density decreases, the nitrogen density remains practically constant. At 10~ppb $\mbox{O}_2$ a streamer still emerged. At a lower purity of $\mbox{O}_2$, 1~ppb, we run into the limitations of the fluid approximation used in our model, with less than one oxygen molecule per cell at the finest level. Initially, a very thin streamer slowly emerged with a very high electric field (600~kV/cm). However, after decreasing the computational domain and increasing the spatial resolution of the simulation, the streamer did not emerge anymore. In addition, the original simulation showed some artifacts near the symmetry axis. In the original simulation, cell-sizes ranged from 1.39~$\mu$m (finest level) to 11.1~$\mu$m (coarsest level). The follow-up simulation used cell-sizes between 0.347~$\mu$m and 5.56~$\mu$m. Both results can be seen in figure~\ref{fig:n2_1ppb}. Since the low density of $\mbox{O}_2$ makes the applicability of the fluid approximation questionable, we can't make any claims about the possibility of photo-ionization as a mechanism for positive streamer propagation at these levels of purity. Additionally, experimentally testing nitrogen with such a high purity would require large effort and investments.

\section{Comparison with experiments}

\subsection{Velocity, diameter and branching}

The present investigations were inspired by experiments conducted by Nijdam \textit{et al.} \cite{Nijdam:2010,Nijdam:2009} on streamers in gas compositions similar to the ones we used in the numerical simulations. The experiments produce pictures of the optical emissions of streamers that typically branch repeatedly. Nevertheless, they can be compared qualitatively with the numerical results. Figure 7 in \cite{Nijdam:2010} shows a comparison between streamers in air and streamers in pure nitrogen. The pure nitrogen has a contamination of $\mbox{O}_2$ of less than 1~ppm. The streamers in air are about twice as thick as those in nitrogen. Both propagate with the same velocity.

Streamer initiation and propagation has also been observed in pure oxygen (less than 10~ppm contamination). However, this only occurred at higher voltages. For a given gas pressure, roughly twice the voltage was needed compared to the other nitrogen-oxygen mixtures. Unfortunately, streamers in pure oxygen emit very little light and are therefore very difficult to image and analyze \cite{Nijdam:2010}. Nonetheless, it seems that their general morphology, diameter and propagation velocity is not far off from the other gas mixtures. In addition, positive streamers with similar velocities and diameters were also observed in argon, while in argon properties such as branching behavior and light emission are different and the streamers emerge more easily at lower voltages than in N$_2$:O$_2$ mixtures.

\begin{figure}
\center{\includegraphics[width=0.5\columnwidth]{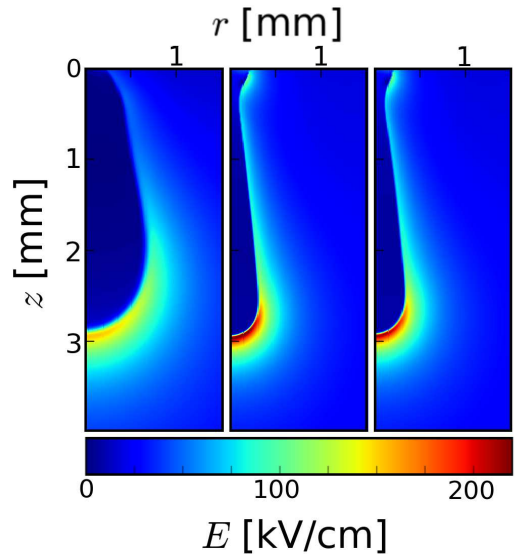}}
\caption{Electric field strength of three streamers of equal length in an 8~mm gap. The left figure is in air with photo-ionization and no background ionization, the middle figure is in $\mbox{N}_2$ with 1ppm $\mbox{O}_2$, photo-ionization and no background ionization. The right figure is $\mbox{N}_2$ with 1ppm $\mbox{O}_2$ with $10^7$ cm$^{-3}$ background ionization and no photo-ionization. The computational domain is larger than the plotted area.}\label{fig:airp_n2p_n2i}
\end{figure}
Figure~\ref{fig:airp_n2p_n2i} shows similar results from the numerical simulations: The streamer in air is thicker than its nitrogen counterparts. Unlike the situation in air, the streamer generated by photo-ionization and the one generated by background ionization look remarkably similar, though it is important to note that both pictures are not from the same time step, they have been selected so that each streamer is in the same stage of its propagation and the streamer generated by photo-ionization is slightly faster, as was discussed earlier. The data used is the same as in figure~\ref{fig:n2p_n2i_airp_elec_dens}. Note that both streamers in nitrogen started to branch not long after the timestep shown in the figure. Due to the cylindrical symmetry of the system, the simulations were stopped after the streamer branched.

\begin{figure}
\center{\includegraphics[width=0.5\columnwidth]{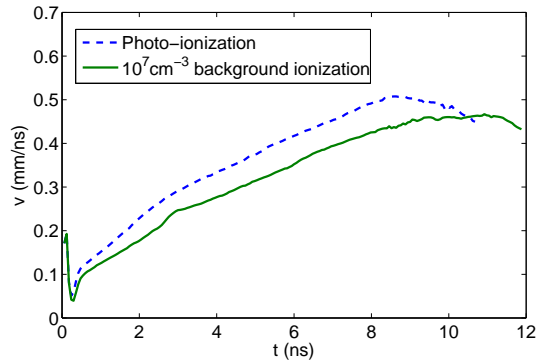}}
\caption{Propagation velocities of streamers in nitrogen using different propagation mechanisms.}\label{fig:n2_velocities}
\end{figure}
The propagation velocity of streamers in $\mbox{N}_2$ is not constant, but increases slowly up to the branching point. In Figure~\ref{fig:n2_velocities} we see that the maximum velocity reached is approximately $0.5$~mm/ns. In experiments conducted by Nijdam \textit{et al.} \cite{Nijdam:2010} in a 16~cm gap at 200~mbar a linear relation was found between voltage and the velocity halfway down the discharge gap. At 20~kV, they measured a velocity of $0.4$~mm/ns. We must note that we can not mimic the conditions of the experiments precisely, as we can not follow the streamer after it has branched. The same experiments show that streamers in nitrogen branch more easily than in air. In our simulations, branching was observed in nitrogen, but not in air under otherwise identical conditions.

\subsection{Feather-like structures}

In their experiments in ``pure'' nitrogen and argon, Nijdam \textit{et al.} \cite{Nijdam:2010} observed feather-like structures on the sides of streamer channels. This was not observed in mixtures with 1\% or more oxygen, in which streamer channels appear smooth. They hypothesized that these feathers are electron avalanches generated by single electrons that move into the region where the electric field is above the breakdown threshold. A distinction must be made between nitrogen and air: in nitrogen, the photo-ionization length scale is much larger than the distance from the streamer head at which the electric field exceeds the breakdown threshold. Therefore many of the electrons that are created by photo-ionization do not form avalanches. Only when an electron reaches the area in front of the streamer head, an avalanche will form. This results in a low number of avalanches that can be seen as distinct feathers.

In air on other hand, the photo-ionization length is much smaller and electrons created by photo-ionization are immediately accelerated to form an avalanche in the area of high electric field. This results in a lot of avalanches that are no longer distinct, but will instead overlap and become part of the streamer head that can become much wider for precisely this reason. This heuristic argument matches the observation that streamer channels in air are straight and wide.

In experiments in pure nitrogen at 200~mbar, the visible hairs of the feathers have a length of 0.5--1.5~mm. The angle between these hairs and the propagation direction of the streamer is 20 to 50 degrees. This value is determined from a 2D projection of the real hairs, therefore the real angles may be larger than 20 degrees. It is not clear whether the hairs bend away from the streamer channel or towards it. The maximum distance between the tip of a hair and the center of a streamer channel is about 1~mm, but most hairs do not stick out more than 0.5~mm from the axis of the channel.

About 1--1.5 hairs per mm of streamer channel can be observed. Because hairs in the path of the channel will be overrun by the channel and hairs in the same optical path as the streamer channel will be obscured by the streamer, the total number of avalanches per mm is larger than 1.5 hairs per mm.

Simulating the creation of these feathers goes beyond the capabilities of our fluid model and requires a model that tracks the individual electrons or a spatially hybrid model, such as \cite{Li:2009}, that combines the fluid approximation in the interior of the streamer with the full particle model outside of the streamer. However, we can still make some qualitative statements based on our simulation data.

There are 2 parameters that influence the presence of distinct feathers: the distance from the streamer at which avalanches are created and the number of avalanches per unit volume. If avalanches only occur close to the streamer, they will be indistinguishable from the main streamer channel. Similarly, if the electron density is high enough that they can be described as a density rather than as a probability, one can expect the number of avalanches to be so high that individual avalanches overlap and distinct feathers are no longer visible, but rater one wide channel is seen.

Due to the similarity laws \cite{Briels:2008b,Ebert:2010}, the feather length of 1.0~mm seen in the 200~mbar experiments in nitrogen corresponds to a length of 0.2~mm in our simulations at 1000~mbar. In our simulations in air with photo-ionization, we found the electron density to be at around $10^5$~mm$^{-3}$ at distances of 1~mm (or more) from the streamer channel. Therefore we expect any effects from individual avalanches to be smoothed out and to be invisible as the density-description holds. This is in agreement with experiments, where no distinct feathers were seen in air.

In nitrogen with photoionization the situation is different, the electron density falls off more rapidly with distance and we observe that at 0.2~mm from the streamer head, the electron density drops to 10$^2$~mm$^{-3}$, which is sufficiently low to consider the electron density distribution as a probability distribution rather than as a continuous medium. We note that at 0.2~mm to the side of the streamer head, the electric field is around 80~kV/cm, well over the breakdown threshold, so avalanches should be able to start at this distance and even further from the streamer. So in nitrogen we find that we both have a sufficiently low electron density as well as a sufficiently high electric field to enable the formation of distinct avalanches at at least 0.2~mm from the streamer head.

In conclusion, we again emphasize that a proper investigation of these feather-like structures is not possible with our fluid model and requires a particle model. However, our arguments based on the density of electrons qualitatively agree with the observed differences between the smooth streamers in air and the feathered streamers in nitrogen.

\section{Conclusion}

We have simulated and analyzed the propagation of positive streamers due to photo-ionization or background ionization, in air as well as in nitrogen with 1 ppm oxygen which corresponds to the lowest impurity level reached in experiments~\cite{Nijdam:2010}. In such pure gases, the usual photo-ionization mechanism present in air is largely suppressed. The initial background ionization can come from natural radioactive sources or from residual ionization from a previous discharge.

We have found that in air the photo-ionization mechanism dominates the streamer propagation except when a very high density of background ionization, such as 10$^{10}$ cm$^{-3}$, is present. (This background density can be associated with a repetition frequency of 1~kHz according to the estimates in section 2.2.) We have found that the parameters of the photo-ionization model have a very small effect, relative to the change in number of ionizing photons, on the streamer characteristics: an order of magnitude change of the number of ionizing photons results in a change of 20 \% in streamer characteristics such as the velocity. Therefore we conclude that although the detailed parameters of the photo-ionization model are not well known, we still expect that the numerical results will hold up experimentally.

In nitrogen with 1 ppm oxygen, we found that photo-ionization is still dominating streamer propagation up to background ionization levels of 10$^{7}$ cm$^{-3}$ (corresponding to a repetition frequency of 1~Hz). This is remarkable since the low oxygen concentration leads to a low number of photo-ionization events per volume. As lower impurity levels than 1 ppm are extremely difficult to reach experimentally, we conclude that streamer propagation even in "pure" nitrogen is dominated by the usual photo-ionization mechanism in non-repetitive discharges.

While for all simulations with photo-ionization or background ionization in different gas compositions, the streamer velocity changes by less than a factor of two, there are characteristic differences in shape and field enhancement. The nonlocal photo-ionization in air creates a wide electron cloud around the streamer head that can be interpreted as a density; this explains why the streamer head in air can become broad and propagate in a stable manner. On the other hand, pure background ionization in air or in nitrogen as well as the weak photo-ionization in nitrogen with 1 ppm oxygen create a steep decrease of the electron density around the streamer head. These densities become so low immediately outside the streamer head that they have to be interpreted as probabilities rather than as densities, hence creating a more stochastic propagation mode in which the streamer cannot become as wide as in air. These observations match the experiments \cite{Nijdam:2010} that show a more feathery structure consisting of many avalanches around thin streamer channels in "pure" nitrogen while streamers in air are straighter and wider. Velocities are comparable between air and "pure" nitrogen both in experiments and in our simulations.

The simulations show another characteristic difference between streamers in air and in "pure" nitrogen that up to now cannot be verified in experiments: The field enhancement at the streamer tip is stronger in nitrogen than in air. This is reminiscent of the difference between positive and negative streamers in air. Negative streamers in air become wider along the channel due to electron drift \cite{Luque:2008} and at their head due to the non-local photo-ionization; therefore they are not very able to keep the field focussed. For positive streamers in air, the field focussing is suppressed at their head through photo-ionization, while positive streamers in "pure" nitrogen stay narrow and focus the field at the heads to the highest values. Therefore they create higher ionization levels in the streamer channel, and they can propagate with similar velocities as in air though the electron density falls off faster ahead of the ionization front.

We have studied photo-ionization versus background ionization for positive streamers in air and in "pure" nitrogen. We showed that for sufficiently low repetition frequencies and background ionization, photo-ionization is dominant in both gases, but that streamers can propagate by pure background ionization as well, and in a similar manner. We discussed characteristic differences of propagation modes between strong or weak photo-ionization. Finally, we believe that our results are representative for other gas composition as well.

\ack
G.W. acknowledges support by STW-project 10118, part of the Netherlands' Organization for Scientific Research (NWO). A.L. was supported by the Spanish Ministry of Science and Innovation under project AYA2009-14027-C05-02.

\section*{References}

\bibliographystyle{iopart-num}
\bibliography{Wormeester2010_JPD}

\end{document}